\documentclass[twocolumn,tighten,times]{aastex63}
\usepackage{warning}  
\usepackage{threeparttable}
\usepackage{longtable}

\usepackage{amsmath}
\usepackage{xcolor}  
\usepackage{soul}  
\allowdisplaybreaks

\renewcommand{\deg}{$^{\circ}$}

\newcommand{\Rom}[1]{\uppercase\expandafter{\romannumeral #1\relax}}

\newcommand{\Rsun}{$R_{\Sun}$}

\journalinfo{This Preprint originally submitted to AAS Journals, November 5, 2019 and accepted December 11, 2019. See published version for peer-reviewed Version of Record.}

\shorttitle{Multi-line single point coronal magnetometry}
\shortauthors{Dima \& Schad}

\begin{document}

\title{\Large{Using multi-line spectropolarimetric observations of forbidden emission lines to measure single-point coronal magnetic fields}}

  
\correspondingauthor{Gabriel I Dima}
\email{gdima@nso.edu}

\author[0000-0002-6003-4646]{Gabriel I. Dima}
\affil{National Solar Observatory \\
22 Ohia Ku St. \\
Pukalani, Hawaii 96768, USA}

\author[0000-0002-7451-9804]{Thomas A. Schad}
\affil{National Solar Observatory \\
22 Ohia Ku St. \\
Pukalani, Hawaii 96768, USA}

\begin{abstract}
Polarized magnetic dipole (M1) emission lines provide important diagnostics for the magnetic field dominating the evolution of the solar corona.  This paper advances a multi-line technique using specific combinations of M1 lines to infer the full vector magnetic field for regions of optically thin emission that can be localized along a given line of sight.  Our analytical formalism is a generalization of the ``single-point inversion" approach introduced by \citet{Plowman14}. We show that combinations of M1 transitions for which each is either a $J=1\rightarrow0$ transition or has equal Land\'e g-factors for the upper and lower levels contain degenerate spectropolarimetric information that prohibits the application of the single-point inversion technique.  This may include the pair of \ion{Fe}{13} lines discussed by \citet{Plowman14}.  We identify the \ion{Fe}{13} 10747 \AA{} and \ion{Si}{10} 14301 \AA{} lines as one  alternative combination for implementing this technique.  Our sensitivity analysis, based on coronal loop properties, suggests that for photon noise levels around $10^{-4}$ of the line intensity, which will be achievable with the National Science Foundation's \textit{Daniel K. Inouye Solar Telescope}, magnetic fields with sufficient strength (${\sim}10$ G) and not severely inclined to the line-of-sight ($\lesssim 35^{\circ}$) can be recovered with this method.  Degenerate solutions exist; though, we discuss how added constraints may help resolve them or reduce their number.  
\end{abstract}


\section{Introduction} \label{sec:intro}
In the low density environment of the solar corona, the magnetic field governs both the atmospheric structure and the energetics of active solar phenomena like flares, coronal mass ejections and the solar wind.  Two promising techniques available for remotely measuring the coronal magnetic field are spectropolarimetric measurements of emission lines in the optical part of the spectrum and radio observations from coronal regions with different plasma frequencies \citep{Casini17}; however, each face observational and interpretation challenges.

The brightest observable visible/infrared (Vis/IR) coronal emission lines are magnetic dipole (M1) lines \citep{Judge98, DelZanna18a}. Photoexcitation of these lines by anisotropic disc radiation leads to scattering induced linear polarization with amplitudes of ${\sim}10^{-2}$ to $10^{-1}$ of the total line intensity \citep[e.g.,][]{Arnaud87, Tomczyk08}). The presence of the magnetic field modifies the linear polarization through the saturated Hanle effect (\textit{i.e.}, its amplitude is not sensitive to the magnetic field intensity) and induces a small circularly polarized signal through the longitudinal Zeeman effect with relative amplitudes of ${\sim}10^{-4}$ to $10^{-3}$ for a 10 G magnetic field \citep{Harvey69, Lin00, Lin04}. Achieving high sensitivity measurements of these signals is challenged by the large ${\sim}10^{5}$ brightness contrast between the solar disc and coronal emission in the Vis/IR \citep{Arnaud82, Arnaud87, Kuhn96, Madsen19} and requires mitigation of scattered light from the sky and telescope optics. Observations at longer wavelengths are typically more sensitive because the scattered light performance of telescope optics improves and the Zeeman splitting increases faster than the thermal line width \citep{Penn14}. The upcoming 4m \textit{Daniel K. Inouye Solar Telescope} (DKIST) will have two first light spectropolarimetric instruments targeting multiple IR emission lines built to address these challenges: DL-NIRSP and Cryo-NIRSP \citep{Rimmele15}.

Key theoretical advances in understanding the formation of polarized M1 lines in the corona have enabled new diagnostic capabilities based on synthesizing polarimetric signals through the optically thin corona \citep{Charvin65,SahalBrechot77,House77,Casini99,Casini00}.  Due in part to the current sparsity of circular polarization detections, most progress has been made by comparing observed linear polarized signals of a single M1 line to those synthesized through forward models \citep{Judge06,Dove11,Rachmeler13,Dalmasse19} or by using time-based tomographic methods to invert the global coronal magnetic structure \citep{Kramar06, Kramar13}, again primarily using a single M1 line.  These techniques necessarily consider the emission contributed by all the plasma along each line of sight.  Work by \cite{Liu08} succeeded in comparing measured circular polarization signals with those synthesized through an active region.  

In contrast to the above approaches, here we focus on using full Stokes observations of multiple M1 lines to invert the full vector magnetic field for an isolated region with uniform magnetohydrodynamic properties: density, temperature and magnetic field. This region either dominates the total emission along the line of sight or can potentially be separated from the foreground/background emission with subtraction methods. Good candidate targets are coronal loops that appear in contrast against the background corona in ultraviolet \citep{Brooks12,Aschwanden13,Xie17} and IR \citep{Tomczyk08}. \citet{Plowman14} demonstrated that, under this ``single point approximation'', using a combination of full Stokes observations from the \ion{Fe}{13} 10747 \AA{} and 10798 \AA{} lines, all three components of the magnetic field could be analytically calculated without explicit knowledge of the thermodynamic parameters that influence the amplitude of the linear polarization. 

In this work we examine the single point inversion method further by generalizing the approach of \citet{Plowman14} to consider, in principle, any combination of M1 lines.  The analytical formalism we derive in Section \ref{analytic_solution} for two line observations reveals that only specific line combinations may be used for the inversion due to fundamental quantum mechanical degeneracies in the polarized emission coefficients. As a result, we learn that the \ion{Fe}{13} line pair used by \citet{Plowman14} may not be adequate for single point inversions if LS coupling is valid for these lines.  In Section \ref{sec:noise} we propose combining the \ion{Fe}{13} 10747 \AA{} and \ion{Si}{10} 14301 \AA{} coronal lines to implement the single point inversion.  We also evaluate the impact photon noise has on this line combination when attempting to infer the magnetic field vector.   In Section \ref{discussion} we discuss ambiguities inherent in the present inversion approach and as well as other related diagnostics.

\section{Analytic inverse solution}
\label{analytic_solution}

\begin{figure}
    \centering
    \includegraphics[width=3.3in]{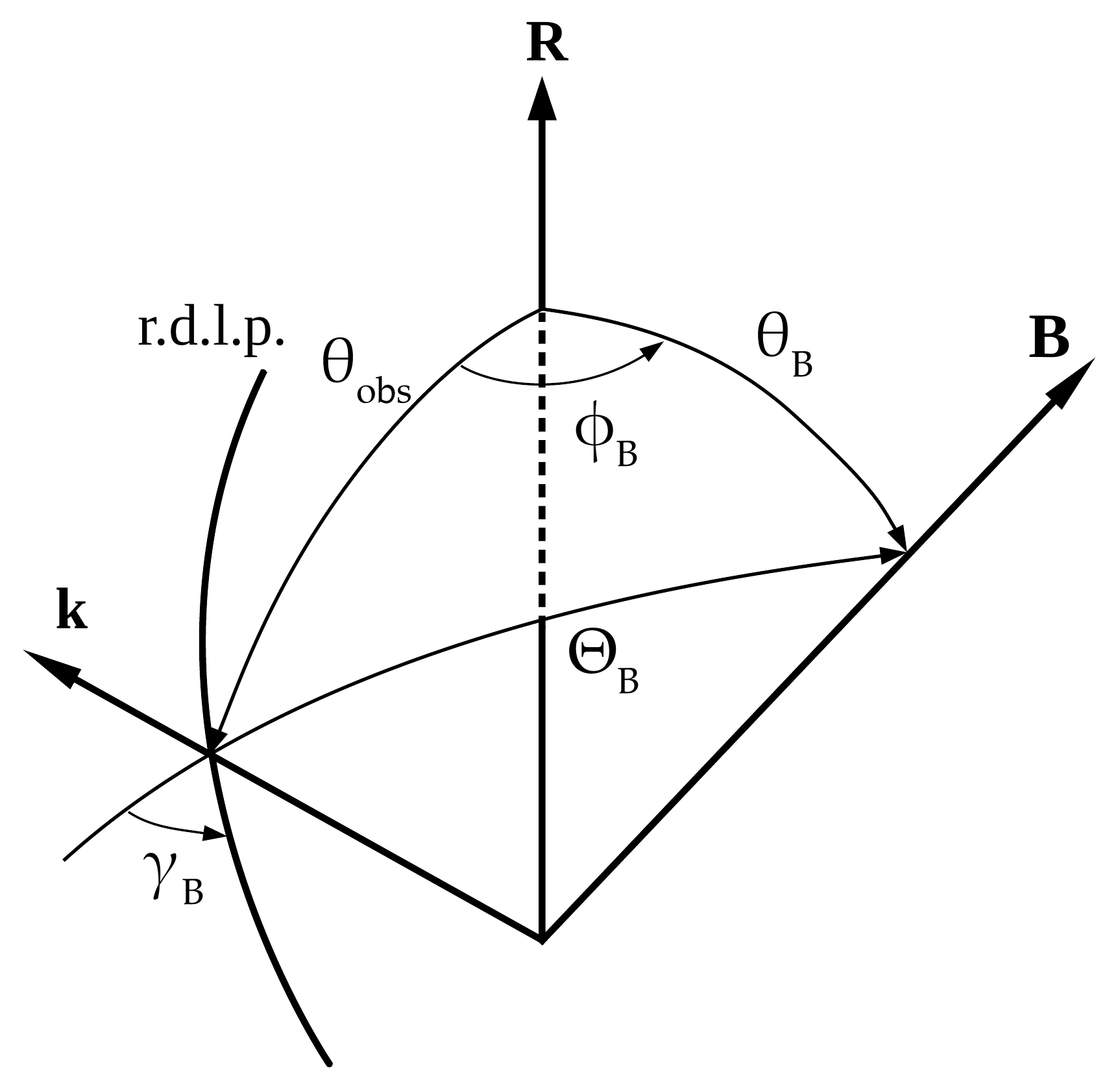}
    \caption{Angular coordinates between three vectors passing through a point of emission in the corona: the outward radial direction (\textbf{R}), magnetic field (\textbf{B}) and line of sight (\textbf{k}). The projected magnetic field in a plane perpendicular to the line of sight is characterized by the azimuth angle $\gamma_B$ measured clockwise towards the reference direction for linear polarization (r.d.l.p.) with a 180\deg ambiguity since the linear polarization direction is not oriented. Figure adapted from Figure 5 in \citet{Casini99}.} 
    \label{fig:coordinate_system}
\end{figure}

Starting from equations 35(a-c) in \citet{Casini99,Casini00}, and making use of the geometric tensors for magnetic dipole radiation given by their equations 39(a-d), the polarized emission coefficients ($\epsilon_{i}$) for the Stokes parameters ($i=I,Q,U,V$) of an M1 line can be written as a function of wavelength (as opposed to angular frequency) as:
\begin{align}
  \epsilon_I(\lambda) &= C\left[1 + \frac{3}{2\sqrt{2}}D\sigma^2_0\left(\cos^2{\Theta_B} - \frac{1}{3}\right)\right]\phi(\lambda-\lambda_0)\label{eqI0}\\
  \epsilon_Q(\lambda) &= \frac{3}{2\sqrt{2}}CD\sigma^2_0\sin^2{\Theta_B}\cos{(2\gamma_B)}\phi(\lambda-\lambda_0) \label{eqQ0}\\
  \epsilon_U(\lambda) &= -\frac{3}{2\sqrt{2}}CD\sigma^2_0\sin^2{\Theta_B}\sin{(2\gamma_B)}\phi(\lambda-\lambda_0) \label{eqU0}\\
  \epsilon_V(\lambda) &= -C\frac{\lambda^2}{c}\nu_L\cos{\Theta_B}\left[\bar{g}+E\sigma^2_0\right]\phi'(\lambda-\lambda_0) \label{eqV0}
\end{align}
where $C$ (\textit{i.e.}, $C_{JJ_{0}}$ in the original notation) is the emission coefficient (without stimulated emission) in the unpolarized case and depends on the total upper level population and the Einstein A coefficient for the transition. $D=D_{JJ_{0}}$, $E=E_{JJ_{0}}$ and $\bar{g}=\bar{g}_{JJ_{0}}$ in the original notation are atomic parameters that depend on the quantum numbers of the upper and lower levels of the transition, $J$ and $J_0$ respectively. $\Theta_B$ is the inclination angle of the magnetic field with respect to the line of sight while $\gamma_B$ is the azimuth angle of the reference direction of linear polarization (r.d.l.p) measured from the projected magnetic field direction (see Figure \ref{fig:coordinate_system} as well as Figure 5 in \cite{Casini99}). The Larmor frequency $\nu_{L}$ is proportional to the magnetic field strength B. $\phi(\lambda-\lambda_0)$ refers to the Doppler dominated normalized ($\int \phi{\ }d\lambda = 1$) spectral profile centered at $\lambda_0$ while $\phi'(\lambda-\lambda_0)$ is its derivative with respect to wavelength. 

The scalar quantity $\sigma^2_0$ refers to the fractional atomic alignment of the upper level of the transition which describes the imbalance of the magnetic sublevel populations for different absolute values of the magnetic quantum number.  It is equal to $\rho^2_0/\rho^0_0$ where the tensor $\rho^{K}_{Q}$ is the irreducible spherical statistical tensor representation of the atomic density matrix.  $\sigma^2_0$ can be positive or negative and is generally a function of the exciting photospheric radiation anisotropy (which varies with height), the plasma density and temperature, and the local inclination angle of the magnetic field in the local radial frame of reference $\theta_B$.  For the M1 lines of interest, $\sigma^2_0$ can often be well approximated by a positive definite factor, $k_{J}(T_{e},n_{e},h),$ dependent on the electron temperature, electron density, and height multiplied by a functional dependence on $\theta_B$ given by $3 \cos^2{\theta_B} - 1$ \citep{Judge07}, which gives rise to the Van Vleck ambiguity discussed in Section~\ref{ambiguities}.  Note that $\sigma^2_0$ does not depend on the magnetic field intensity.  

In principle, given proper constraints on the local plasma properties and the radiation field, as well as numerical calculations of $k_{J}$, full Stokes observations of a single M1 line can be used to derive the full magnetic field vector, within certain limitations and within the single point approximation as discussed in \cite{Judge07}.  In absence of such extra constraints, a full set of Stokes observations of a single M1 line provide four measurements for five unknowns (\textit{i.e.}, $C,\sigma^2_0,\gamma_B, \Theta_{B},B$).  $\gamma_B$ can be derived using the ratio of U and Q; but the remaining variables remain coupled.  With some limitations, to be discussed in Section~\ref{blos}, the projected magnetic field intensity along the line of sight ($B_{LOS}\equiv B\cos\Theta_B$) can be derived from polarimetric observations of a single line. Likewise, the value of $C$ can be determined from I by neglecting the atomic alignment contribution; but $\sigma^2_0$ and $\Theta_B$ cannot be directly determined.

\citet{Plowman14} showed that with full Stokes measurements of two M1 lines, in particular the \ion{Fe}{13} 10747, 10798 \AA{} line pair, it is possible to obtain explicit analytic formulas for all the unknown parameters in equations \ref{eqI0}-\ref{eqV0} under the single point assumption.  An additional spectral line provides four more measured quantities but only two additional unknowns, as the magnetic field properties are assumed to be the same.  \citeauthor{Plowman14} focused only on the \ion{Fe}{13} lines, using analytic formulas for the D and E parameters from Equation 38(a-b) in \citet{Casini99} before solving for the unknown parameters. 

To extend this approach to other combinations of M1 emission lines, here we retain the symbolic form for the parameters D, E. Under the single point approximation, the line is strongly dominated by emission from a single region in space so the measured polarized Stokes profiles $S_i(\lambda)$ can be approximated as:
\begin{align}
\begin{array}{ll}
S_i(\lambda) = \int_0^{\infty}\epsilon_i(\lambda)ds \approx \epsilon_i(\lambda)\Delta s & (i=I,Q,U,V)
\end{array}
\end{align}
where the integral is over emission along the line of sight and $\Delta s$ is the length of the homogeneous single point region. From the measured dispersed Stokes profiles the wavelength-independent Stokes amplitudes for the two lines can then be derived as: 
\begin{align}
  I &= \frac{S_I}{\phi(\lambda-\lambda_{0})} \approx \frac{\epsilon_I\Delta s}{\phi(\lambda-\lambda_{0})} \nonumber \\
  &\approx I_0\left[1 + \frac{3}{2\sqrt{2}}D\sigma^2_0\left(\cos^2{\Theta_B} - \frac{1}{3}\right)\right] \label{eqI}\\
  L &= \frac{\sqrt{S_Q^2 + S_U^2}}{\phi(\lambda-\lambda_{0})} \approx \frac{\sqrt{\epsilon_Q^2 + \epsilon_U^2}\Delta s}{\phi(\lambda-\lambda_{0})} \nonumber \\
  &\approx \pm \frac{3}{2\sqrt{2}} D I_0 \sigma^2_0 \sin^2{\Theta_B} \label{eqL}\\
  V &= \frac{cS_V}{\lambda^2\phi'(\lambda-\lambda_{0})} \approx \frac{c\epsilon_V\Delta s}{\lambda^2\phi'(\lambda-\lambda_{0})} \nonumber \\
  &\approx -\frac{1}{\sqrt{2}}I_0\nu_L\cos{\Theta_B}\left[\left(\sqrt{2}+D\sigma^2_0\right)\bar{g}+FD\sigma^2_0\right] \label{eqV}
\end{align}
 where $I_0 \equiv C\Delta s$ is a constant with units of intensity. We combine the Q and U Stokes parameters into a single parameter L, which is the total linear polarization.  We keep both signs of the quadratic sum because, as already stated, the sign of the atomic alignment $\sigma^2_0$ can be either positive or negative while the measured quantity L is always positive. We also introduce the parameter F defined as:
\begin{align}
  F &\equiv \sqrt{2}\frac{E}{D} - \bar{g} \nonumber\\
    &= \frac{3}{4}[J(J+1) - J_0(J_0+1) -2](g_1 - g_0) \label{eqF}
\end{align}
where $g_1$ and $g_0$ are the Land\'e factors for the upper and lower levels in the transition. The analytic expression for F can be derived from Equations 36b-c in \citet{Casini99} for $D$ and $E$ by expanding the Wigner 6-j and 9-j coefficients and is valid for all M1 transitions with $\Delta J=0,\pm1$. This parameter is introduced in hindsight since it will be shown to play an important role in discriminating between lines that contain non-degenerate information. 

The key to the two-line inversion involves solving for the angle $\Theta_B$ that gives the same value for B assuming lines are emitted from the same spatial point with the same magnetic field properties.  We first write the magnetic field strength B as a function of $\Theta_B$ by substituting equations \ref{eqI} and \ref{eqL} into equation \ref{eqV} and using the expression for the Larmor frequency (\textit{i.e.}, $\nu_L=(\mu_B/h)B$):
\begin{align}
  B = \frac{h}{\mu_B}\frac{-V}{\left[(I \pm L)\bar{g} \pm \frac{2}{3}F\frac{L}{\sin^2{\Theta_B}}\right]\cos{\Theta_B}} \label{eqB_num1}
\end{align}
where $h$ and the $\mu_B$ are Planck's constant and the Bohr magneton respectively.  Using this equation and subscripts 1 and 2 for the two lines, we assume $B_{1} \equiv B_{2}$ and $\Theta_{B1} \equiv \Theta_{B2}$ and then solve for $\sin^{2}\Theta_{B}$.  The other unknown parameters then follow from Equations~\ref{eqI},~\ref{eqL},and~\ref{eqB_num1}.  To summarize, the analytical inversion equations are as follows: 
\begin{align}
  \sin^{2}\Theta_{B} &= \pm \frac{2}{3}\frac{F_1L_1V_2 - F_2L_2V_1}{\bar{g}_2V_1(I_2 \pm L_2) - \bar{g}_1V_2(I_1 \pm L_1)}, \label{eqsin}\\[5pt]
  I_{0i} &= I_i \pm L_i\left(1 - \frac{2}{3\sin^2{\Theta_B}}\right), \\[5pt]
  \sigma^{2}_{0i} &= \frac{\pm L_i}{\sqrt{\frac{9}{8}} I_{0i} D_i \sin^{2}\Theta_{B}}, \\[5pt]
  B &= \frac{h}{\mu_B}\frac{-V_{i}}{\left[(I_i \pm L_i)\bar{g}_i \pm \frac{2}{3}F_i\frac{L_i}{\sin^{2}{\Theta_{B}}}\right]\cos{\Theta_{B}}},\label{eqB}\\
  \gamma_B &= \left\{
  \begin{array}{ll}
    -\frac{1}{2}\arctan\left(\frac{U_i}{Q_i}\right) + \{0,\pi\}, & \sigma^2_{0i} > 0 \\[5pt]
    -\frac{1}{2}\arctan\left(\frac{U_i}{Q_i}\right) + \{\frac{\pi}{2}, \frac{3\pi}{2}\}, & \sigma^2_{0i} < 0 \\
  \end{array}
  \right. \label{eqgamma}
\end{align}
where the arctangent function must account for the signs of the arguments to determine the quadrant for the result (usually implemented numerically as the arctan2 function). Since M1 lines are in the saturated Hanle regime, the magnetic field projection ($\gamma_B$) is parallel or perpendicular to the direction of linear polarization depending on sign of the alignment and the field inclination relative to the outward radial direction $\theta_B$ (Van Vleck effect, see discussion in Section \ref{ambiguities}). In addition, the reference direction of linear polarization has a $180^{\circ}$ ambiguity, which introduces the additional solutions indicated in the curly brackets of Equation~\ref{eqgamma}.  Consequently, given two measured Stokes vectors this inversion method produces up to four degenerate solutions for the magnetic field (discussed further in Sections~\ref{sec:noise} and ~\ref{ambiguities}).  The sign for $\cos\Theta_B$ is chosen such that B is positive in equation \ref{eqB} since physically the magnetic field strength is a positive quantity.  

An important result can be seen from equation \ref{eqsin}: emission lines with $F=0$ contain degenerate information about the magnetic field such that no unique solution for $\Theta_B$ can be deduced from the two Stokes vectors. To apply this type of inversion emission lines must be selected such that at least one line has $F\ne0$.  From Equation~\ref{eqF}, F is zero when the upper and lower levels have equal Land\'e g-factors ($g_1=g_0$) or for $J=1\rightarrow0$ transitions.  

Table~\ref{tab:lines} reports the values of the D and F atomic parameters for a list of candidate visible and infrared M1 lines identified by \citet{Judge01} as having the best coronagraphic potential.  The values are calculated assuming LS coupling such that the Land\'e factors for each level are given by
\begin{align}
  g = \frac{3}{2} + \frac{S(S+1) - L(L+1)}{2J(J+1)},\label{eq:LScoupling}
\end{align}
where S and L are the total spin and orbital angular momenta.  We find that only three of the candidate lines have $F\ne0$---\ion{Fe}{10} 5303 \AA{}, \ion{Si}{10} 14301 \AA, and \ion{Mg}{8} 30285 \AA. The \ion{Fe}{10} 6375 \AA{} line can only be circularly polarized because the upper level has $J=1/2$ and it therefore cannot be atomically aligned. Transitions with an upper level that is not polarizable have $D=0$, $F=0$ and no alignment correction is needed to measure an unbiased value for the line of sight magnetic field (see discussion in Section \ref{blos}).

Based on these results, the \ion{Fe}{13} 10747 \AA{} and 10798 \AA{} lines considered by \citet{Plowman14} each have $F=0$ and thus contain degenerate information.  Consequently, this line pair does not allow one to invert for the magnetic field vector. The discrepancy between his results and ours arises due to the use of Equation 38b from \citet{Casini99} which contains a small algebraic error leading to a spurious value for the coefficient $E$ for the \ion{Fe}{13} 10798 \AA{} line.\footnote{We have confirmed this error with the authors of \cite{Casini99} and \cite{Plowman14}.}  That said, the information in these line pairs is only degenerate if LS coupling holds for the energy levels.  This may not be case for the \ion{Fe}{13} 10747, 10798 \AA{} pair because the separation in the energy levels for the two transitions does not obey Land\'e's interval rule, which states that the splitting between two fine structure levels of the same term is proportional to the larger J value. The ratio of energies between the $^3P_2 \rightarrow\ ^3P_1$ and $^3P_1 \rightarrow\ ^3P_0$ transitions in the ground term of \ion{Fe}{13} is 9258.6/9303.1 $\sim1$ instead of 2 as would be expected from Land\'e's interval rule. It may be that the \ion{Fe}{13} line pair does contain non-degenerate information but confirmation of the Land\'e factors for the levels is needed before the line can be used as a diagnostic for the magnetic field. 

Recent laboratory experiments by \citet{Hensel16} and \citet{Bekker18}, using electron beam ion traps, succeeded in measuring the Land\'e g-factors for visible transitions from highly ionized iron, specifically \ion{Fe}{14} 5303 \AA, \ion{Fe}{10} 6475 \AA{} and \ion{Fe}{11} 7892 \AA. The g-factors measured for \ion{Fe}{11} 7892 \AA{} indicate significant departures from LS coupling; however, the g-factors for the two levels of the transition are equal within the experimental uncertainties which still means $F=0$ for this line.  Extending such laboratory measurements to the infrared transitions of iron, as well as the other species in Table \ref{tab:lines}, is necessary to ensure the accuracy of coronal magnetic field measurements. 

\begin{deluxetable}{lr|ccccccccccccc}
\tablecaption{D and F atomic parameters for candidate visible and IR magnetic dipole coronal lines selected by \cite{Judge01} assuming LS coupling for the Land\'e g-factors. \label{tab:lines}}
\tablehead{
  \colhead{Ion} & \colhead{$\lambda_0$(air)[\AA]} & \colhead{Transition (u$\rightarrow$l)} &
  \colhead{D$_{JJ_0}$} & \colhead{F$_{JJ_0}$}
}
\startdata
\ion{Fe}{14} & 5303\tablenotemark{a} & $^2P_{3/2} \rightarrow\ ^2P_{1/2}$ & 0.71 & 0.5\\
\ion{Fe}{10} & 6375\tablenotemark{a} & $^2P_{1/2} \rightarrow\ ^2P_{3/2}$ & 0.0 & -\tablenotemark{e}\\
\ion{Fe}{11} & 7892\tablenotemark{a} & $^3P_1 \rightarrow\ ^3P_2$ & 0.11 & 0.0 \\
\ion{Fe}{13} & 10746\tablenotemark{b} & $^3P_1 \rightarrow\ ^3P_0$  & 1.0 &  0.0\\
\ion{Fe}{13} & 10798\tablenotemark{c} & $^3P_2 \rightarrow\ ^3P_1$ & 0.6 & 0.0\\
\ion{S}{9}   & 12524\tablenotemark{d} & $^3P_1 \rightarrow\ ^3P_2$ & 0.1 & 0.0 \\
\ion{Si}{10} & 14301\tablenotemark{b} & $^2P_{3/2} \rightarrow\ ^2P_{1/2}$ & 0.71 & 0.5\\
\ion{S}{11}  & 19201\tablenotemark{d} & $^3P_1 \rightarrow\ ^3P_0$ & 1.0 & 0.0 \\
\ion{Si}{7} & 24826\tablenotemark{d}  & $^3P_2 \rightarrow\ ^3P_1$ & 0.1 & 0.0 \\
\ion{Mg}{8} & 30285\tablenotemark{d} & $^2P_{3/2} \rightarrow\ ^2P_{1/2}$ & 0.71 & 0.5\\
\ion{Si}{9} & 39343\tablenotemark{c}  & $^3P_1 \rightarrow\ ^3P_0$ & 1.0 & 0.0 \\
\ion{Mg}{7} & 55033\tablenotemark{d}  & $^3P_2 \rightarrow\ ^3P_1$ & 0.6 & 0.0\\
\ion{Mg}{7} & 90090\tablenotemark{d}  & $^3P_1 \rightarrow\ ^3P_0$ & 1.0 & 0.0
\enddata
\tablenotetext{a}{Laboratory measurements of g-factors available from \citet{Bekker18} but not used here.}
\tablenotetext{b}{\citet{Dima19a}}
\tablenotetext{c}{\citet{DelZanna18a}}
\tablenotetext{d}{CHIANTI version 8 \citep{DelZanna15}}
\tablenotetext{e}{Not defined since $D{=}0$ and $E{=}0$}
\end{deluxetable}

\section{The influence of observational noise \\ on inversions of \texorpdfstring{\ion{Fe}{13} and \ion{Si}{10}}{Fe XIII and Si X }}\label{sec:noise}

Coronal measurements are background-limited \citep{Penn04}, and current observatories like COMP \citep{Tomczyk08} can only routinely achieve noise levels at the $10^{-3}$ noise level, while upcoming facilities like DKIST \citep{Rimmele15} and COSMO \citep{Tomczyk16} will more routinely achieve observations at the $10^{-4}$ noise level and higher.  Following \citet{Plowman14}, here we consider the influence of observational noise on the inverted quantities using a line combination that meets the criteria set forth above. 

We propose the \ion{Fe}{13} 10747 \AA{} ($F=0$) and \ion{Si}{10} 14301 \AA{} ($F=0.5$) line combination as a promising alternative to the \ion{Fe}{13} line pair.  Of the lines with $F\ne0$ identified, the \ion{Si}{10} 14301 \AA{} has perhaps the best potential for full Stokes measurements on account of its brightness and heightened Zeeman sensitivity due to its longer wavelength \citep{Judge01}.  Meanwhile \ion{Fe}{13} 10747 \AA{} has already been demonstrated to be a line with good potential for coronal magnetometry \citep{Lin00,Lin04}.  Both lines are photoexcited from the lower level of the ground state and have large values for D, which in turn leads to larger atomic alignments and larger linear polarization.  While the temperatures of peak ionization fraction for \ion{Fe}{13} and \ion{Si}{10} are different (${\sim}$1.7 MK and 1.4 MK, respectively), the contribution functions do overlap significantly such that structures like coronal loops between 1.5 and 1.6 MK will have measurable emission in both lines.  Furthermore, both lines have now been measured simultaneously during eclipses and using ground-based coronagraphic linear polarimetry \citep{Dima18, Dima19a}. One drawback to using line pairs from different ionized species is the increased likelihood that emission in each line may not originate in the same coronal region. However, as discussed in Section \ref{sec:implementation} analyzing the polarization angle for each line offers an important check on whether the emission is co-spatial.

\begin{figure}
    \centering
    \includegraphics[width=3.3in]{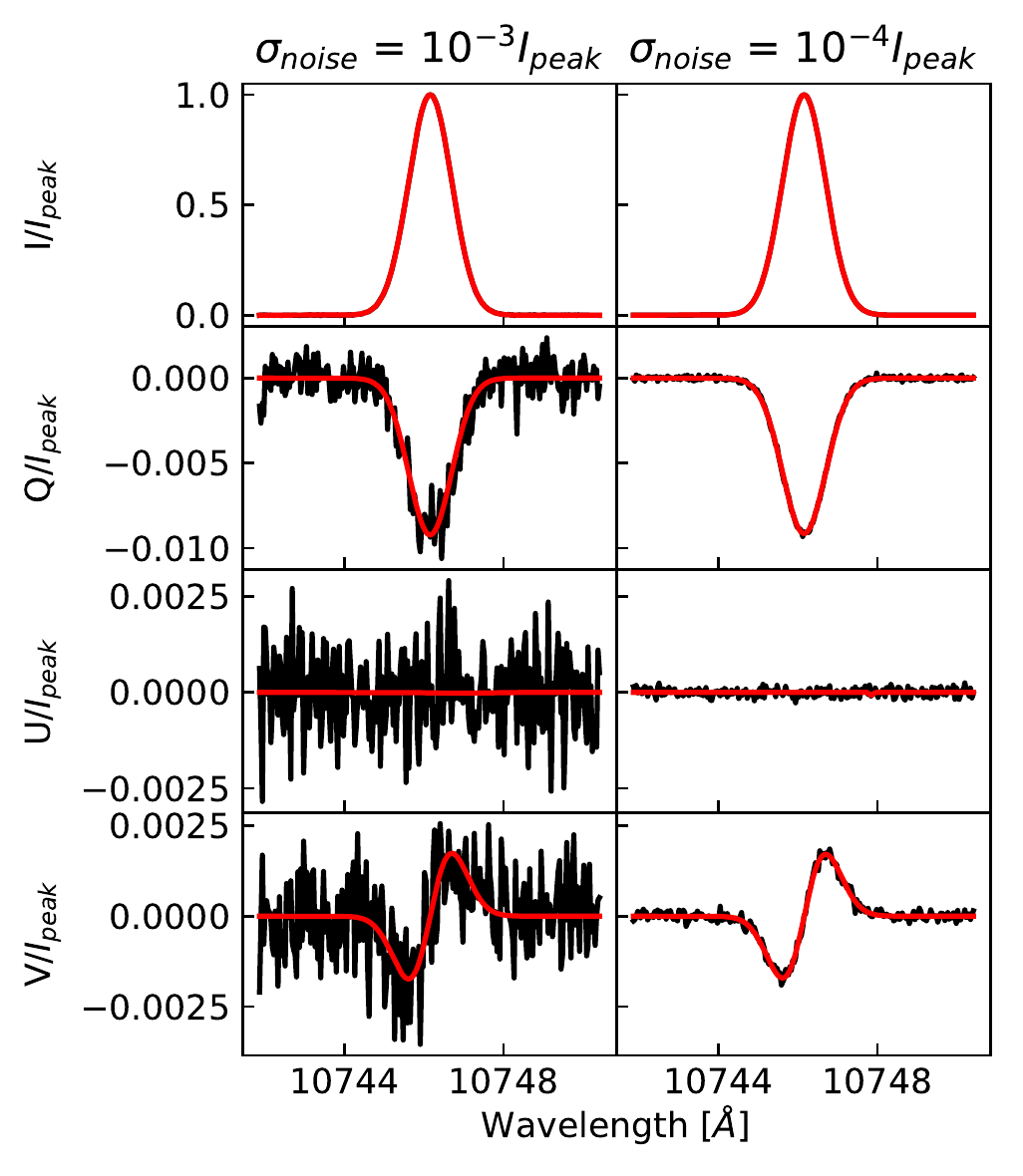}
    \caption{Synthetic noisy line profiles (black) and fits (red) for the \ion{Fe}{13} 10747 \AA{} line. The spectra sampling is 0.035 \AA{} and normally distributed random photon noise is added to each pixel with the standard deviation indicated above each column. The emission region is located at a height of 0.1 \Rsun{} and the scattering angle is $\theta_B = 90^{\circ}$. The input magnetic field has line of sight coordinates (B, $\Theta_B$, $\gamma_B$) = (25 G, 40\deg, 0\deg).} 
    \label{fig:line_fit_noise}
\end{figure}

Here we investigate the effect of noise on inversions of this line combination by performing a Monte Carlo experiment based on inverting synthetic observations of the two lines.  Using a modified version of the polarized coronal line emission code \citep[CLE,][]{Judge01b}, which numerically calculates the atomic alignment and level populations for statistical equilibrium of a multi-level atom including photoexcitation and electron collisions, we generated dispersed full Stokes line profiles for the \ion{Fe}{13} 10747 \AA{} and \ion{Si}{10} 14301 \AA{} lines sampled at 0.035 \AA.  Example line profiles of the \ion{Fe}{13} 10747 \AA{} line are shown in Figure \ref{fig:line_fit_noise} for normally distributed noise in each pixel with standard deviations of $10^{-3}I_{peak}$ and $10^{-4}I_{peak}$.  The \ion{Si}{10} profiles are comparable and not shown.  The spectra shown are for emission located in the plane of the sky passing through the center of the Sun and at a height of $0.1$ \Rsun{} above the photosphere.  The reference direction for linear polarization is perpendicular to the plane containing the line-of-sight and the radial direction. The input magnetic field has local radial coordinates (B, $\theta_B$, $\phi_B$) = (25 G, 90\deg, 40\deg) and line of sight coordinates (B, $\Theta_B$, $\gamma_B$) = (25 G, 40\deg, 0\deg).  The electron density is set to $n_e = 10^{8.5}$ cm$^{-3}$ and the temperature is $\sim1.55$ MK.  As we are using relative noise values, we do not need to set a value for the total emission measure (or length of plasma along line of sight); however, the general scenario here is that of an over-dense closed coronal loop observed near its apex above the solar limb \citep[see, \textit{e.g.}][]{Aschwanden13}. 

Like \cite{Plowman14} we test the single point inversion for different inclination angles relative to the line of sight ($\Theta_B$) for constant values of the atomic alignments ($\sigma^2_0$).  To achieve this we fix the strength of the field B and $\theta_{B}$ while varying the azimuth $\phi_B$ in the plane perpendicular to the local radial direction (see Figure \ref{fig:coordinate_system}), which in turn varies $\Theta_B$.  Physically this is akin to changing the orientation of a coronal loop from an edge-on ($\Theta_{B} = 0^{\circ}$) to a face-on ($\Theta_{B} = 90^{\circ}$) observational geometry. This approach, in addition to fixing $B \equiv 25$ G, is comparable to that of \cite{Plowman14}.  Note, however, that the errors derived will depend upon the magnetic field intensity and the atomic alignments; here only one representative example is considered.  

\begin{figure*}
    \centering
    \includegraphics[width=6.5in]{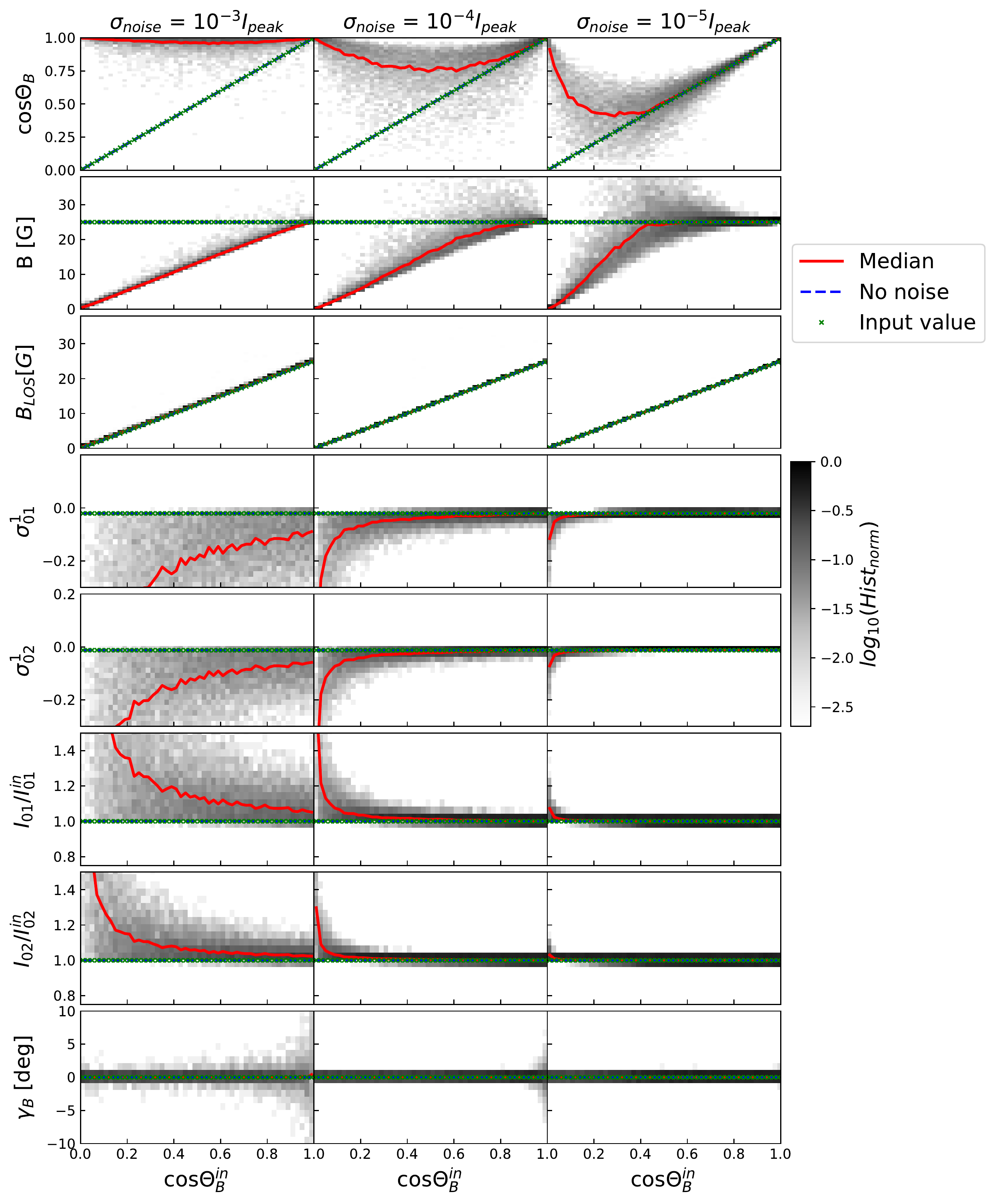}
    \caption{Normalized histograms for the parameters calculated using equations \ref{eqsin} - \ref{eqgamma} choosing the negative solution branch for different levels of noise. Each parameter is shown as a function of the input inclination with respect to the line of sight ($\Theta^{in}_B$). Since the original solution had a local inclination angle $\theta_B = 90^{\circ}$ which is larger than the Van Vleck angle the input atomic alignment is negative and this solution branch accurately recovers the input parameters in the noise-free case.} 
    \label{fig:negative_sol}
\end{figure*}

\begin{figure*}
    \centering
    \includegraphics[width=6.5in]{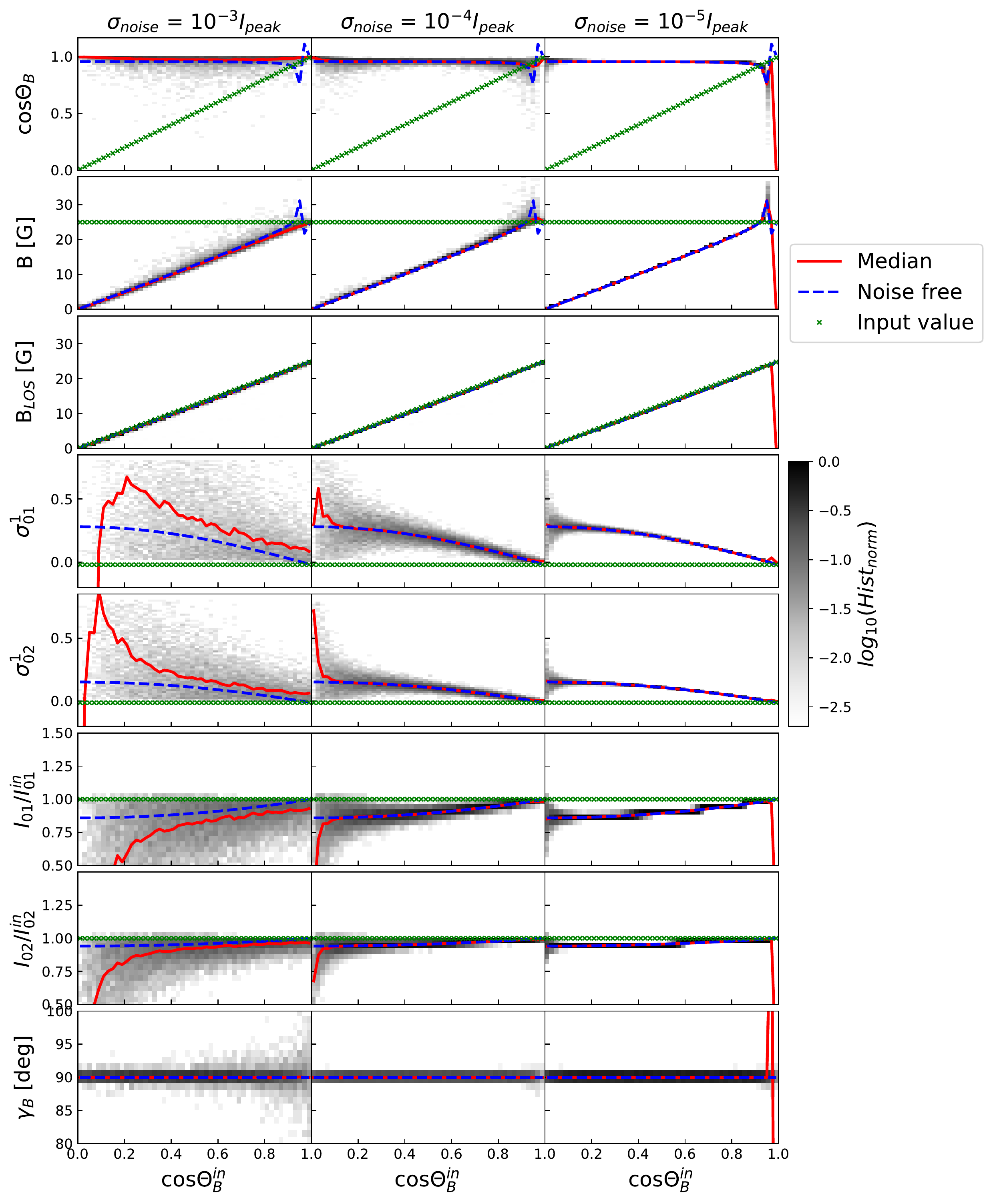}
    \caption{Same as figure \ref{fig:negative_sol} but using the positive solution branch. At very low inclination angles ($\Theta_B \lesssim 17^{\circ}$) this solution branch gives inconsistent solutions ($\cos\Theta_B > 1$) effectively solving the $90^{\circ}$ ambiguity. As discussed in the text regions where this happens depend on both the lines used and the physical properties of the emission region.
    It is also noticeable that at larger inclination angles the recovered atomic alignments $\sigma^2_{0i}$ are very large. Atomic modeling can be used to rule out solutions with such large atomic alignments based on the scattering geometry.} 
    \label{fig:positive_sol}
\end{figure*}

For values of $\Theta^{in}_B$ ranging from $0^{\circ}$ to $90^{\circ}$, where the added superscript indicates ``input'' values, we produce a sample of 500 profiles with different realization of random noise.  For each profile, we determine the Stokes amplitude values of I, L, and V in Equations~\ref{eqI}-\ref{eqV} by fitting in a least-squares sense for the normalized profile $\phi(\lambda-\lambda_{0})$ and its derivative $\phi'(\lambda-\lambda_{0})$.  The inverted parameters at each noise instance are then obtained from Equations \ref{eqsin}-\ref{eqgamma} and the solutions for each parameter are binned into histograms at each input $\Theta^{in}_B$. $I_{01}$,$\sigma^2_{01}$ and $I_{02}$,$\sigma^2_{02}$ refer to the \ion{Fe}{13} and \ion{Si}{10} lines, respectively.  The input values for alignments are $\sigma^2_{01}=-0.021$ and $\sigma^2_{02}=-0.012$.  The normalized histograms for the negative and positive solution branches of Equations \ref{eqsin}-\ref{eqgamma} are shown in Figures \ref{fig:negative_sol} and \ref{fig:positive_sol}.  Note that $\gamma_B$ and $\gamma_B + \pi$ are also solutions in either case.

In this particular example the input atomic alignments are negative and correct solutions for the parameters are recovered by the negative solution branch (Figure \ref{fig:negative_sol}). The positive solution branch shown in Figure \ref{fig:positive_sol} recovers formally consistent solutions (i.e. solutions with $\mid\cos\Theta_B\mid \leq 1$) for $\Theta^{in}_B > 17^{\circ}$ while for $\Theta^{in}_B<17^{\circ}$ the inversions produce inconsistent solutions with $\cos \Theta_B > 1$. In general the range of $\Theta_B$ for which the number of degenerate solutions decreases is a function of both the lines analyzed and the atomic alignments as discussed in Section \ref{ambiguities}.

We find similar noise effects on the inverted parameters using the \ion{Fe}{13}, \ion{Si}{10} line combination to results discussed by \citet{Plowman14}. Noise levels of $\sim10^{-4}I_{peak}$ ($\sim10^{-5}I_{peak}$) are needed to recover input magnetic field inclinations $\Theta^{in}_B < 36^{\circ}$ ( $\Theta^{in}_B < 66^{\circ}$). Parameters like the atomic alignments and the line intensities are recovered well for noise levels at the $10^{-4}I_{peak}$ levels. The line of sight magnetic field, $B_{LOS}$, and projected magnetic field angle, $\gamma_B$, are particular robust to noise. 

In general the full magnetic field vector can be recovered for field inclinations within some limiting angle to the line of sight, $\Theta_B < \Theta^{lim}_B$. The value $\Theta^{lim}_B$ depends on the quantum mechanical properties of the lines used in the inversion, the atomic alignments and the observation noise levels. In the example considered here with a moderately strong field $B=25$ G and small values for the atomic alignments, $\Theta^{lim}_B \approx 36^{\circ}$ for $\sim10^{-4}I_{peak}$ noise levels. Larger noise levels, smaller values for B and smaller atomic alignments all lead to a decrease in $\Theta^{lim}_B$ and a shrinking of the range of invertible solutions.

In agreement with \citet{Plowman14}, we note that the inverted $\Theta_B$ value is biased towards $0^{\circ}$ for larger relative noise and for more transverse input fields directions.  \citeauthor{Plowman14} alleviated this bias by applying a Bayesian MCMC method and appropriate priors to an observation and mapping the posterior probability distribution at each input value of $\Theta_B$. Such a method could be applied to any line pair assuming, as demonstrated, at least one of the lines has $F\ne0$. 

\section{Discussion}\label{discussion}

\subsection{Ambiguous Solutions}\label{ambiguities}
\begin{figure}
    \centering
    \includegraphics[width=3.3in]{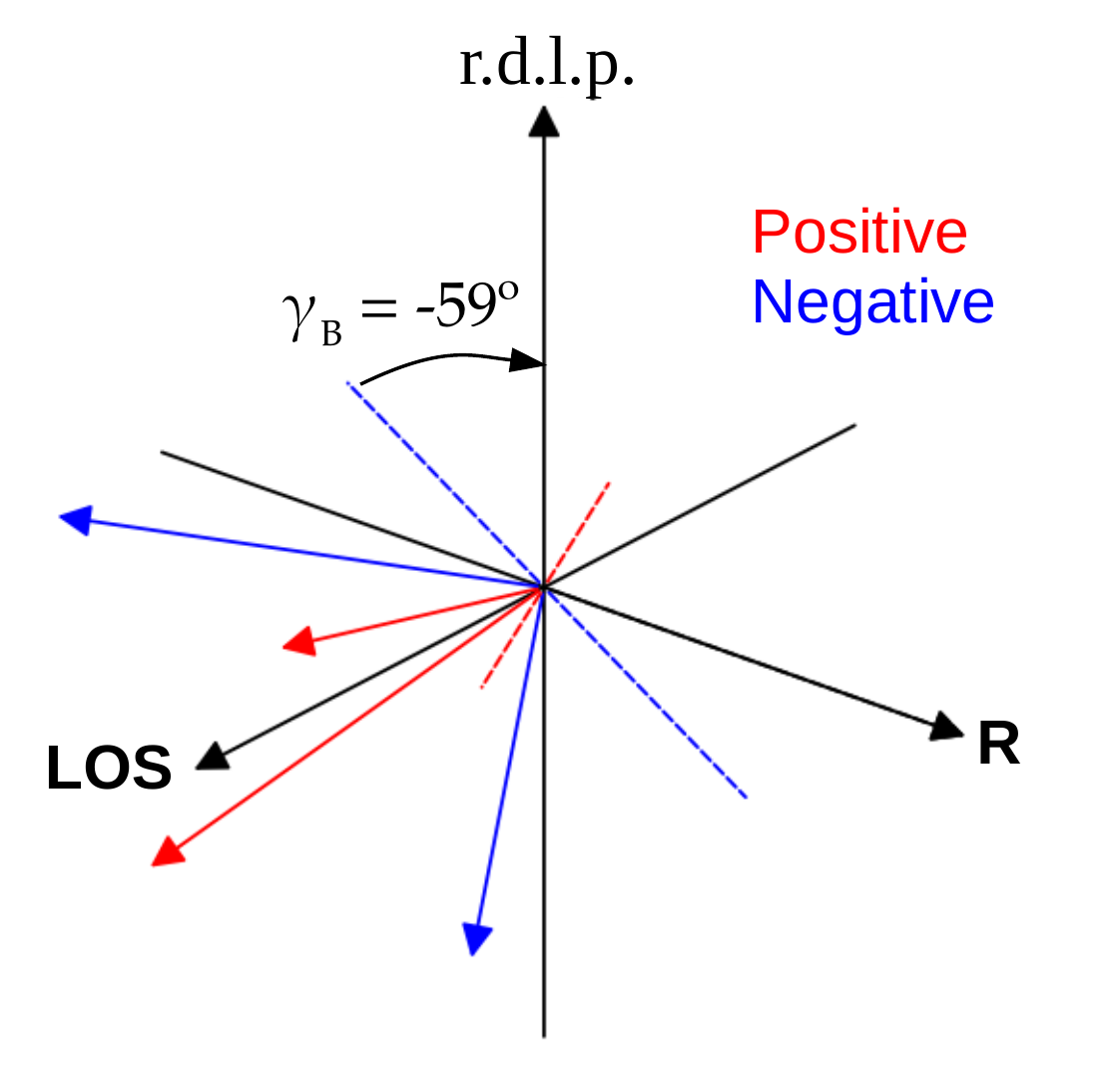}
    \caption{Plot of four degenerate solutions shown in the LOS frame of reference. Scattering angle is $\theta_{obs} = 90^{\circ}$ so the emission is located in the plane perpendicular to the LOS and bisecting the Sun.
    The input field is (B, $\Theta_B$, $\gamma_B$) = (25 G, 35\deg, -59\deg) and corresponds one of the arrows shown in blue. The other blue arrow corresponds to the 180\deg ambiguous solution due to the lack of orientation in the reference direction for the linear polarization (r.d.l.p). The red pair of arrows represents the degenerate solutions due to the Van Vleck ambiguity. The dashed lines represent the projections of the solutions on the plane perpendicular to the line of sight.} 
    \label{fig:all_solutions}
\end{figure}

In the most general case equations \ref{eqsin}-\ref{eqgamma} return four consistent solutions ($\mid\cos\Theta_B\mid \leq 1$) for the inverted set of parameters. Two of the solutions have projected directions $\gamma_B$ at 90\deg{} to each other and map to the classical Van Vleck ambiguity. However, it is notable from looking at Figures \ref{fig:negative_sol} and \ref{fig:positive_sol} that the ambiguity extends to all the inverted parameters. Thus solving this ambiguity may be accomplished by imposing physical limits on any of the solution parameters. The other two ambiguous solutions have parameters identical to the first two except the projected angles $\gamma_B$ are different by 180\deg{} (see example in Figure \ref{fig:all_solutions}). 

The number of solutions depends on both the intrinsic quantum mechanical properties of the transitions through the $D$, $F$ and $\bar{g}$ parameters and on the atomic alignments $\sigma^2_{0i}$ which are functions of the extrinsic thermodynamic properties of the plasma and radiation anisotropy. An example of this is discussed in Section \ref{sec:noise}, where the number of solutions decreases to only two (180\deg{} ambiguity) when $\Theta_B\lesssim 17^{\circ}$. When the emission lines used in the inversion have equal $D$ and $F$ values (e.g. \ion{Fe}{14} 5303 \AA{} and \ion{Si}{10} 14301 \AA{} or \ion{Mg}{8} 30285 \AA) the number of degenerate solutions, in the noise free case, reduces to two for all values of $\Theta_B$ and atomic alignments.

While the multi-line method recovers (at most) four possible degenerate solutions for the vector magnetic field and atomic alignment, it does not provide direct knowledge of the scattering angle $\theta_{obs}$, the location of the emission along the line of sight, nor the magnetic field direction in a coordinate frame local to the Sun.  The field direction in the solar frame can be calculated from the inferred (degenerate) LOS-frame field vector(s) for a given scattering angle using spherical trigonometry (see equations 42,44 in \citet{Casini99}).  With knowledge of the projected height above the solar surface and the inferred values of atomic alignment, numerical calculations (as discussed above) can determine a range of temperature and densities required to be consistent with the observations at a given scattering angle.  Consequently, independent observations of the temperature and density can constrain the scattering geometry itself and the field in the local frame as discussed in the next section.  See also further discussion by \citet{Judge07} and \citet{Schad18}.   

Even without knowledge of the plasma density and temperature, however, some of the degenerate solutions may be ruled out for particular scattering angles based on the relationship between the sign of $\sigma^2_0$ and $\theta_B$, given by the ($3 \cos^2{\theta_B} - 1$) term discussed above \citep{Judge07}. At $\theta_{B} \approx 54.74^{\circ}$ (and ${\approx}125.26^{\circ}$), referred to as the Van Vleck angle, the atomic alignment changes sign. At each scattering angle along the line of sight, the solution pairs ($\sigma^2_0$, $\theta_B$) may be valid or determined to be invalid based on the sign relationship, or as discussed in \citet{Schad18}, ruled out if the scattering angle is unphysically large given the observations. 

\subsection{Combining multi-line M1 inversions with other plasma diagnostics}

\begin{figure}
    \centering
    \includegraphics[width=3.3in]{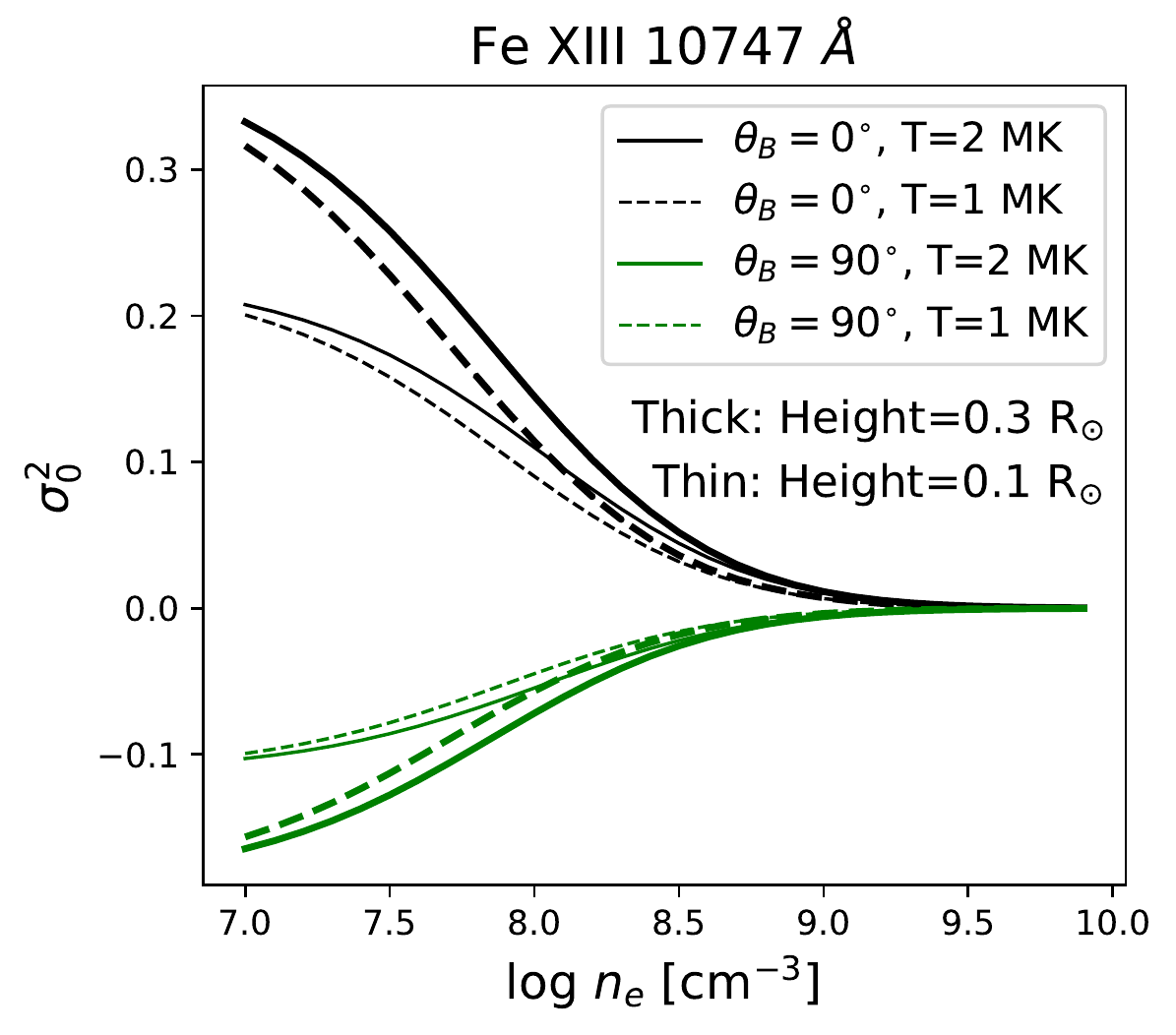}
    \caption{Variation of the atomic alignment for the \ion{Fe}{13} 10747 \AA{} line with density, temperature and magnetic field inclination $\theta_B$ with respect to the incident radiation axis of symmetry. The atomic alignment is calculated at a height of 0.1 \Rsun (thin lines) and 0.3 \Rsun (thick lines).} 
    \label{fig:sigma_density}
\end{figure}

Combining multi-line M1 inversions with spectroscopic plasma diagnostics for density and temperature provides an opportunity to further constrain the ambiguity of the M1 inversion, as discussed by \cite{Judge13}, as well as a means to cross-validate numerical calculations of the atomic alignment with model independent empirical determinations. In addition, it is anticipated a simultaneous model fit of two or more M1 lines (as long as at least one of the lines has $F\ne0$) will further improve resilience of the solution to observational noise. Once again, the atomic alignment depends on electron density ($n_e$) and temperature ($T_e$), height above the photosphere and the magnetic field orientation relative to the local radial direction $\theta_B$ (see Figure \ref{fig:sigma_density}). Emission line ratios of both ultraviolet and infrared transitions may be used as plasma diagnostics to independently measure $n_e$ and $T_e$ \citep[e.g. see reviews by][]{DelZanna18a,delZanna18b}. This information can be used, in addition to knowledge of the projected height above the limb, to determine the scattering angle consistent with the multi-line M1 inverted parameters.  Separately, the atomic alignment can be numerically calculated given the inferred plasma parameters, geometry, and field direction and therefore compared with the inferred value of $\sigma^2_0$ as a way to benchmark the theoretical calculations.  However, we note that very high signal-to-noise is required to precisely determine the alignment (see Figure~\ref{fig:negative_sol}). 

\subsection{A single line measurement of \texorpdfstring{$B_{LOS}$ at a single point}{B LOS at a single point}}\label{blos}

While so far we have focused on multi-line diagnostics, the formalism we presented in Section \ref{analytic_solution} also helps identify diagnostic techniques available for single M1 line spectropolarimetric observations.  In particular, it has been known since \cite{Casini99} that the atomic alignment plays a role in the circular polarized amplitude and that the typical weak field magnetograph formula cannot be directly applied to measure the longitudinal component of magnetic field (\textit{i.e.} along the line-of-sight (LOS)) without an associated scale error.  Interestingly, \cite{Plowman14} demonstrated via his Equation 14 that the scale error could by mitigated by using the observed linear polarized amplitude as a correction.  In our generalization of this approach, the updated form of this relation is given via Equation \ref{eqB} as     
\begin{align}
  B^{obs}_{LOS} &= \frac{h}{\mu_B}\frac{-V}{\left[(I \pm L)\bar{g} \pm \frac{2}{3}F\frac{L}{\sin^{2}{\Theta_{B}}}\right]}. \label{eqBLOS}
\end{align}
where $B^{obs}_{LOS}$ is the observationally determined longitudinal magnetic field strength.  

\begin{figure}
    \centering
    \includegraphics[width=3.3in]{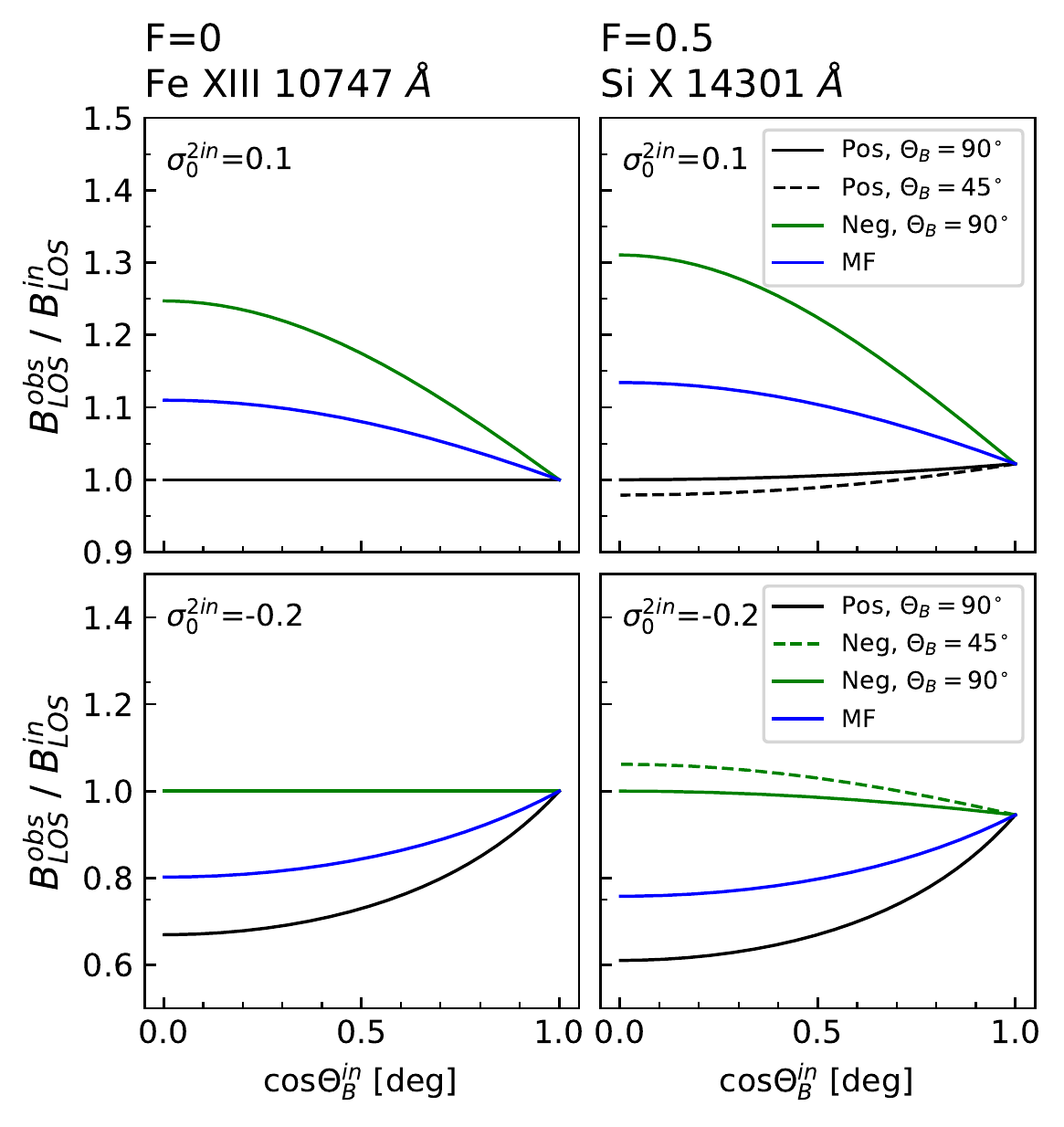}
    \caption{Angle dependent ratios of the measured (or ``observed'') line of sight magnetic field $B^{obs}_{LOS}$ to the true input magnetic field $B^{in}_{LOS}$. The panels on the left correspond to measurements for a line with $F=0$ while the panels on the right correspond to a line with $F=0.5$, with the top (bottom) panels having positive (negative) input atomic alignments $\sigma^{2in}_0$. The black (green) curves represent $B^{obs}_{LOS}$ values calculated with the positive (negative) sign in equation \ref{eqBLOS} and having set $\Theta^g_B=90^{\circ}$ which minimizes the bias at all values for $\Theta_B$. For comparison the dashed lines assume $\Theta^g_B=45^{\circ}$. The blue curves represent $B^{obs}_{LOS}$ calculated using the magnetograph formula (MF, equation \ref{eqMF}).} 
    \label{fig:blos_measurement}
\end{figure}

Two solutions for $B^{obs}_{LOS}$ are obtained which correspond to the positive and negative branches of equation \ref{eqBLOS}.  For lines with $F=0$ (left panels in Figure \ref{fig:blos_measurement}), the solution branch that matches the sign of the atomic alignment accurately measures the input magnetic field ($B^{in}_{LOS}$) independent of $\Theta^{in}_B$ and $\sigma^{2in}_0$. The magnetic field measured for the degenerate solution over-predicts the input magnetic field by an amount that increases with both $\Theta^{in}_B$ and $\sigma^{2in}_0$ so fields more tangential to the LOS have a larger bias than fields parallel to the LOS. 
An example of this effect can be observed for the example discussed in Section \ref{sec:noise}.
Note how for the negative solution branch (Figure~\ref{fig:negative_sol}) the inverted value of $B_{LOS}$ correlates much more tightly with the input value than the positive solution branch (Figure~\ref{fig:positive_sol}).  As the F value for \ion{Fe}{13} 10747 \AA{} is zero, V and L measurements provide tight constraints on $B_{LOS}$ via Equation~\ref{eqBLOS} that are independent of $\Theta_{B}$ and track the true value of $B_{LOS}$ when the chosen solution matches the sign of the input alignment, in this case the negative branch. For this example the bias is small since the atomic alignment ($\sigma^2_{01}=-0.021$) is small compared to the alignment values used to produce Figure \ref{fig:blos_measurement}.

Meanwhile, for transitions with $F\ne0$ (right panels in Figure \ref{fig:blos_measurement}), both solutions for $B^{obs}_{LOS}$ are functions of $\Theta^{in}_B$ and $\sigma^{2in}_0$. Therefore, in principle, without a priori knowledge of these parameters both the measured values for $B^{obs}_{LOS}$ will have a bias relative to the true value $B^{in}_{LOS}$. It is possible to show that the bias is minimized at all values of $\Theta^{in}_B$ if a guess value of $\Theta^g_B=90^{\circ}$ is used in equation \ref{eqBLOS}. Other guess values for $\Theta^{g}_B$ will improve agreement locally but induce larger errors at all other angles (e.g. see dashed lines with $\Theta^g_B=45^{\circ}$ in the right panels of Figure \ref{fig:blos_measurement}). As discussed by \citet{Casini99} the weak field approximation or magnetograph formula (MF) relating $B_{LOS}$ to the Stokes measurements:
\begin{align}
  B^{obs}_{LOS,MF} &= \frac{h}{\mu_B\bar{g}}\left(\frac{-V}{I}\right) \label{eqMF}
\end{align}
biases the measurements by an amount that depends on the atomic alignment since it effectively assumes $L=0$. Plots for the value of $B^{obs}_{LOS}$ using equation \ref{eqMF} are shown in blue in Figure \ref{fig:blos_measurement} and the values fall between the two solutions.

\subsection{Approaches for implementation and limits to the application of the single point approximation}
\label{sec:implementation}
A reliable implementation of the single point inversion, using for example observations to be made available by the upcoming DKIST instruments, relies not only on using appropriate line combinations and sufficient single-to-noise but also on techniques to target and isolate regions of emission along the line of sight, which may be aided by observational checks on the validity of the single point approximation to a given observation. 

Coronal loop structures provide one viable target for applying the single point approximation as has been done by many authors who study the plasma properties along loops using spectroscopic and spectro-imaging techniques \citep[e.g.][]{Aschwanden99, Brooks12, Xie17}.  In these studies the importance of background/foreground emission subtraction is well known \citep[see][and references within]{Terzo10}. The intensity contrast of loops observed by TRACE (using the most sensitive filter for each loop) was generally found to be $\sim15-20\%$  \citep{Aschwanden05}.  Warm ($\sim$1-2 MK) active region loops tend to have higher densities compared to diffuse emission surrounding the loops \citep{Brooks12, Brooks19}. As seen in Figure \ref{fig:sigma_density}, the atomic alignment decreases sharply with increases in the density (the temperature dependence is much smaller). Therefore, although loops have increased density and correspondingly larger brightness, the linearly polarized signals may be lower than for the background plasma.  As such, the single-to-noise requirements become more demanding in order to implement a reliable background subtraction technique, such as fitting the cross section of the polarized emission profile across the loop.  To understand how best to achieve this subtraction, synthesizing polarized emission through high resolution magnetohydrodynamic (MHD) simulations of the active corona \citep[e.g.][]{Rempel17} and then testing methods may be helpful.  

Given a particular observation, one test for the validity of the single point approximation is provided by the two independent measurements of projected linear polarization angle.  Since M1 lines are in the saturated Hanle regime, the angles $\gamma_B$ measured with different lines emitted from the same region should be equal. The condition is necessary but not sufficient because the possibility exists that background linear polarization may satisfy this condition while obscuring the signal. Resulting signals that have background subtraction applied should also satisfy this condition.

\section{Summary and outlook}

In this article, we have presented a generalization of the dual-line inversion algorithm for the vector magnetic field, proposed by \citet{Plowman14} for the \ion{Fe}{13} 10747, 10798 \AA{} line pair, to other combinations of magnetic dipole (M1) lines. A number of results that emerged from the analysis:
\begin{enumerate}
\item Assuming emission is concentrated at a single point in the corona it is possible to combine polarized observations from two or more M1 lines to invert for the vector magnetic field. However, only some combinations of M1 lines contain non-degenerate information such that a finite set of degenerate solutions can be calculated. Transitions between levels with equal upper/lower level Land\'e factors or $J=1 \rightarrow 0$ transitions cannot be used without adding another line where these conditions do not hold. 
\item One pair of lines that satisfies this requirement is \ion{Fe}{13} 10747 \AA{} and \ion{Si}{10} 14301 \AA{}. Using atomic modelling of the polarized emission coefficients under realistic physical conditions for a coronal loop, we showed that the inclination of the magnetic field with respect to the line of sight, $\Theta_B$, and the magnetic field strength B can be recovered for photon noise levels $\sim10^{-4}I_{peak}$ if the magnetic field is oriented close to the line of sight.  
\item In general the method produces four degenerate magnetic field solutions in the line of sight frame of reference consistent with the Van Vleck and 180\deg{} ambiguities. However, the number of allowed solutions may be smaller depending on the field geometry and the emission lines selected. Independent measurements of the height of emission (through stereoscopy) can be used together with an atomic model and the measured atomic alignment to determine the density and temperature of the emission region. Conversely, measurements of density and temperature (through coordinated observations with diagnostic UV and IR lines) can be used to constrain the height of emission.
\item The line of sight magnetic field $B_{LOS}$, obtained from polarized measurements of a single M1 line, is robust to noise but can be measured without bias only for M1 lines that have equal Land\'e factors for the upper and lower levels, or $J=1 \rightarrow 0$ transitions. Other M1 lines that do not satisfy these conditions will give a biased measurement value for $B_{LOS}$ that varies based on the inclination of the magnetic field with respect to the line of sight $\Theta_B$. However, this bias can be minimized for all values of $\Theta_B$ to better than 10\% even for large values of the atomic alignment if the inclination is assumed to be $\Theta_B=90^{\circ}$. 
\end{enumerate}

The coronagraphic polarimetric capabilities of the upcoming National Science Foundation's \textit{Daniel K. Inouye Solar Telescope} provides an unprecedented opportunity to observe M1 emission from the solar corona at high resolution and high sensitivity such that multi-line techniques for coronal magnetic field diagnostics, including the techniques developed here, can be put into practice.  

\section{Acknowledgments}

We extend our thanks to Roberto Casini, Phil Judge, and Joe Plowman for providing feedback and comments on this work.  The National Solar Observatory (NSO) is operated by the Association of Universities for Research in Astronomy, Inc. (AURA), under cooperative agreement with the National Science Foundation. 

\bibliography{gdrefs}{}
\bibliographystyle{aasjournal}

\listofchanges

\end{document}